\documentstyle[12pt,fleqn]{article}
\def\beq{\begin{equation}}
\def\eeq{\end{equation}}
\def\bu{{\bf u}}
\def\bv{{\bf v}}
\def\bA{{\bf A}}
\def\6{\langle }
\def\9{\rangle }

\begin{document}

\renewcommand{\thefootnote}{\fnsymbol{footnote}}

\vspace*{10mm}
\begin{center}
{\large {\bf Optimal distinction between non-orthogonal quantum 
states}}\\[15mm]

Asher Peres\footnote{Electronic address: peres@photon.technion.ac.il}
and Daniel R. Terno\footnote{Electronic address:
terno@physics.technion.ac.il} \\[8mm]
{\sl Department of Physics, Technion---Israel Institute of Technology,
32\,000 Haifa, Israel}\\[15mm] \end{center}

\noindent{\bf Abstract.} Given a finite set of linearly independent
quantum states, an observer who examines a single quantum system may
sometimes identify its state with certainty. However, unless these
quantum states are orthogonal, there is a finite probability of failure.
A complete solution is given to the problem of optimal distinction of
three states, having arbitrary prior probabilities and arbitrary
detection values. A generalization to more than three states is
outlined.  \newpage

\noindent{\bf 1. Non-orthogonal quantum signals}\medskip

\noindent Quantum information theory is an emerging science, which
combines two traditional disciplines: quantum mechanics and classical
information theory. This subject has many fascinating potential
applications for the transmission and processing of information, and
yields results that cannot be achieved by classical means. A simple
example is the use of quanta that have been prepared according to one of
a finite set of states as signals for the transmission of information.
The possibility of using non-orthogonal quantum states, which has no
classical analogue, is especially interesting for its potential
applications to cryptography (that is, for communication security)~[1].

An observer, faced with such a set of signals whose prior probabilities
are known, may follow various strategies. The approach favored by
information theorists is to maximize the mutual information that can be
acquired in the detection process~[2]: each event is analyzed in a way
from which it is possible to deduce definite posterior probabilities for
the emission of the various signals, and the observer's aim is to reduce
as much as possible the Shannon entropy of the ensemble of signals. On
the other hand, communication engineers attempt to guess what the signal
actually was, and their aim is to miminize the number of errors~[3].
Cryptographers, whose supply of signals is essentially unlimited but for
whom security is paramount, do not want any error at all, but on the
other hand they are ready to lose some fraction of the signals. The
latter strategy is the one that will be investigated in this article.

The case of just two non-orthogonal signals is quite simple and well
known~[4--6]. Recently, Chefles~[7] investigated the case of $N$
linearly independent signals, and obtained some partial results. In the
following, we give a complete treatment of the case of three signals.
Our method can readily be generalized to a larger number of signals (but
explicit calculations become tedious).

In the next section, we introduce a set of positive operator valued
measures which describe generalized quantum measurements. (These are
more general than the projection valued measures corresponding to the
standard, von~Neumann type of mesurement.) An explicit algorithm is
developed, to ensure the positivity of the required matrices.

Optimization (namely, how to maximize the information gain) is discussed
in Sect.~3. We consider the possibility that the various signals may
have different ``values.'' The information gain is defined as the
expected average of the values of detected signals (this includes the
possibility that some types of signals are never identified). It is then
shown in Sect.~4 that even if a measurement fails to identify with
certainty a signal, it still is usually possible to attribute to the
various signals posterior probabilities, so that the observer acquires
at least some mutual information on the emitted signals. Finally,
Sect.~5 briefly discusses an extension of this work to spaces with more
than three dimensions.\bigskip

\noindent{\bf 2. Positive operator valued measures}\medskip

\noindent Consider, in a 3-dimensional complex vector space, three
linearly independent normalized state vectors, $\bu_1,\ \bu_2$, and
$\bu_3$ (we are using here the standard notation for Euclidean vectors,
as no confusion may arise). These vectors have the physical meaning of
signals, and they are, in general, not orthogonal. They occur with
probabilities $p_1,\ p_2$, and $p_3$, respectively. In each measurement
the observer should either identify with certainty one of these signals,
or get an inconclusive answer (the latter will be labelled 0, meaning
``no answer''). The objective is to design a procedure that minimizes
the probability of the inconclusive answer. More generally, we may
attribute different values $C_j$ to the various outcomes (for example,
rare signals with small $p_j$ may have larger values than frequent
signals), and our aim is to maximize the expected gain of information.

Note that the number of outcomes of the measuring process is larger than
the dimensionality of the vector space. Therefore we need ``generalized
measurements'' that are represented by positive operator valued measures
(POVM)~[8]. Namely, we have to construct four positive semi-definite 
matrices $\bA_j$, that satisfy

\beq \sum_{j=0}^3 \bA_j = {\bf 1}, \eeq
where {\bf 1} is the unit matrix. Three of these matrices correspond to
the three input signals, and the remaining one to an inconclusive
answer. It is easily proved~[2] that optimal $\bA_j$ may be taken as
matrices of rank 1. However, the optimal solution may not be unique, and
higher rank matrices may also be optimal, as we shall see below.

By analogy with the well known solution for the case of two input
vectors~[4--6], let us define three auxiliary (unnormalized) vectors
$\bv_j$ as follows:

\beq \bv_1=\bu_2\times\bu_3, \label{v} \eeq 
and cyclic permutations. We thus have

\beq \6\bu_j,\,\bv_i\9=\delta_{ji}\,[\bu_1\bu_2\bu_3], \eeq
where $[\bu_1\bu_2\bu_3]$ stands for the triple product of the input
vectors (that is, the determinant of their components, in any basis).

We then construct with the $\bv_j$ three POVM matrices, which correspond
to outcomes of experiments that give a definite identification of an
input signal:

\beq \bA_j=k_j\,|\bv_j\9\6\bv_j|, \eeq
where the $k_j$ are non-negative numbers, that still have to be
determined. Indeed, the probability that the $j$-th outcome results
from the $i$-th input is

\beq  P_j=\6\bu_i,\,\bA_j\,\bu_i\9=k_j\,|\6\bu_i,\,\bv_j\9|^2.
\label{Pj}\eeq
This vanishes if $j\neq i$. Therefore, observing the $j$-th outcome
implies that the input was $\bu_j$. This result occurs with probability

\beq P_j=k_j\,\Bigl|[\bu_1\bu_2\bu_3]\Bigr|^2. \eeq 
Note that the input states $\bu_j$ must be linearly independent in order
to unambiguously distinguish any one of them. It will be convenient for
future use to introduce the notation

\beq T=\Bigl|[\bu_1\bu_2\bu_3]\Bigr|^2. \eeq
This can also be written as $T=[\bv_1\bv_2\bv_3]$, or 

\beq T=1+s_{12}s_{23}s_{31}+s_{13}s_{32}s_{21}
 -|s_{12}|^2-|s_{23}|^2-|s_{31}|^2, \eeq
where $s_{ij}=\6\bu_i,\,\bu_j\9$.

Finally, the remaining POVM matrix, which indicates an inconclusive
answer, is given by

\beq \bA_0={\bf1}-\sum_{j=1}^{3} \bA_j.  \eeq
The probability of the inconclusive answer is

\beq P_0=\sum_{j=1}^3 p_j\,\6\bu_j,\,\bA_0\,\bu_j\9
 =1-T\,\sum_{j=1}^3 k_j\,p_j. \label{P0} \eeq 
We naturally want the $k_j$ to be as large as possible, in order to
increase the detection probabilities, but their values are bounded above
by the demand of positivity of $\bA_0$. Recall that the necessary and
sufficient conditions for the positivity of a matrix are the positivity
of all the diagonal elements and diagonal subdeterminants, including the
determinant of the entire matrix:

\beq \det\bA_0\ge0 \label {det}. \eeq
In the present case, this last condition is the decisive one that
actually determines the domain of acceptable values of $k_j$. This is
intuitively seen as follows:  when all $k_j$ vanish, $A_0\equiv\bf1$,
which has only positive eigenvalues. As we gradually increase the $k_j$,
one of the eigenvalues of $\bA_0$ will vanish and then become negative.
When it vanishes, the determinant vanishes too (because it is equal to
the product of eigenvalues), and this gives the boundary of the domain
of legal $k_j$. The surface $\det(\bA_0)=0$ consists of several disjoint
parts. The role of other positivity conditions is to eliminate (in
practice, to confirm the elimination of) the irrelevant parts of that
surface.

Explicitly, the condition $\det(\bA_0)=0$ can be written as

\beq 1-\sum_{j=1}^3|\bv_j|^2\,k_j+T\,(k_1k_2+k_2k_3+k_3k_1)-
  T^2\,k_1k_2k_3=0. \label{A0} \eeq
A simple way of obtaining Eq.~(\theequation) is to choose a basis in our
vector space, such that the vector components are as simple as possible.
Let the first basis vector be $\bu_1$ itself, and the second one be a
linear combination of $\bu_1$ and $\bu_2$, with real coefficients. This
determines the third basis vector, up to a phase. We can choose phases
so that $\bu_3$ has at most one complex coefficient. We thus obtain

\beq \bu_1=(1,0,0),\\ \bu_2=(a_2,b_2,0),\\ \bu_3=(a_3,b_3e^{i\beta},c_3).
\label{bu} \eeq
Recall that all these vectors are normalized. It is now easy to write
$\det(\bA_0)$ explicitly in terms of the parameters in Eq.~(\ref{bu}),
and then to express these parameters in terms of the various vectors.
The resulting surface, $\det(\bA_0)=0$, is sketched in Fig.~1, for the
following choice of parameters:

\beq \bu_1=(1,\;0,\;0),\qquad \bu_2=(0.6,\;0.8,\;0),\qquad
 \bu_3=(0.5,\;0.5+0.5i,\;0.5). \eeq

The surface given by Eq.~(\ref{A0}) intersects each $k_j$ axis at
$k_j=|\bv_j|^{-2}$. Note that, in the first octant, this surface is
everywhere convex. This can be seen as follows. Let us cut it by one of
the planes $k_j={\rm const}$. The intersection is a rectangular
hyperbola with asymptotes parallel to the remaining axes. For example,
if we cut the surface (\ref{A0}) by the plane $k_3={\rm const.}$, the
asymptote $k_1\to\infty$ is explicitly obtained by dividing
Eq.~(\ref{A0}) by $k_1$ and then setting  $k_1\to\infty$. This gives

\beq -|\bv_1|^2+T\,(k_2+k_3)-T^2k_2k_3=0. \eeq
It is then easily seen that for any fixed $k_3$ such that
$0<k_3<|\bv_3|^{-2}$, the resulting $k_2$ is positive. This means that,
in the plane $k_3={\rm const.}$, the asymptote $k_1\to\infty$ cuts the
positive part of the $k_2$ axis. The same result holds for any other
choice of section parallel to one of the coordinate planes. This proves
the convexity of the surface in Fig.~1: all these sections are convex
segments of rectangular hyperbolas.\bigskip

\noindent{\bf 3. Optimization}\medskip

Finally, we are left with the problem of finding the set of $k_j$ that
maximize the information gain. The latter is

\beq G=\sum_j C_j\,P_j=T\,\sum_j C_j\,p_j\,k_j, \eeq 
where $C_j$ is the ``value'' of signal $\bu_j$ and use was made of
Eq.~(\ref{Pj}). Define, for brevity,

\beq B_j=C_j\,p_j. \eeq
All points of the plane

\beq \sum_{j=1}^3B_j\,k_j=G/T, \label{G}\eeq
with $k_j\ge0$, lead to the same information gain $G$, provided that
these points belong to the domain of positivity of $\bA_0$. The largest
value of $G$ can be obtained as follows.

Let us imagine that we start with a plane $\sum B_jk_j=X$, with large
positive $X$, so that there is no contact between that plane and the
relevant part of the surface~(\ref{A0}). As we gradually decrease $X$,
the plane will reach a point where it is tangent to that surface (thanks
to its convexity). This happens at the point where the gradient of the
left hand side of~(\ref{A0}) is parallel to the vector $\{B_j\}$. If the
point of contact lies in the first octant, it gives the optimal
solution. It may happen, however, that at this point of contact one of
the $k_j$ is negative, and therefore that point is not a valid solution.
In that case, we further decrease $X$, until a contact point occurs on
one of the coordinate planes (that is, one of the $k_j$ vanishes), or
even at one of the vertices (two of them vanish).

For example, when all $p_j={1\over3}$, and all $C_j=1$, the optimal
result is obtained when $k_1=2.4189$, $k_2=0$, and $k_{3}=0.6719$. This
result means that we sacrifice the possibility of detecting signal
$\bu_2$ in order to get the lowest probability for the inconclusive
answer, as may be seen from Eq.~(\ref{P0}). In the present case, we
obtain $P_0=0.8386$. On the other hand, if we give
different values to the signals, such as $C_1=0.8$, $C_2=1.2$, and
$C_3=1$, the optimal result is obtained with $k_1=2.083$, $k_2=0.2902$,
and $k_3=0.2129$. The probability to get an inconclusive answer then is
slightly higher: $P_0=0.8626$.\bigskip

\noindent{\bf 4. Inconclusive answers still carry some
information}\medskip

\noindent An inconclusive answer is not completely useless (except in
special, highly symmetric cases). For example, if $\bu_1$ is orthogonal
to $\bu_2$ and $\bu_3$, and these are not orthogonal to each other, then
$\bv_1$ is parallel to $\bu_1$, and $\bv_2$ and $\bv_3$ lie in the
$\bu_2\bu_3$ plane. The $\bA_0$ matrix is of rank~1: $\bA_0=|{\bf
w\9\6w}|$, with {\bf w} in the $\bu_2\bu_3$ plane. In such a case, the
signal $\bu_1$ is always detected with certainty, while an inconclusive
result means: either $\bu_2$ or $\bu_3$ (with known posterior
probabilities, as explained below).

In general, for arbitrary $\bu_j$, the optimal $\bA_0$ is a matrix of
rank~2 which can be written in terms of its eigenvalues and
eigenvectors:

\beq \bA_0=\lambda_m\,|{\bf m\9\6m}|+\lambda_n\,|{\bf n\9\6n}|.\eeq
Each one of the two terms on the right hand side is by itself a
legitimate POVM element, so that there can actually be two distinct
inconclusive outcomes. Let us label them $m$ and $n$.

Suppose that the outcome of a generalized measurement turns out to be
$m$.  The prior probability for that result, if the input was $\bu_j$,
is

\beq P_{mj}=p_j\,\lambda_m\,|\6{\bf m},\,\bu_j\9|^2. \eeq
By Bayes's theorem, the posterior probability for input $\bu_j$ upon
observing output $m$ is~[8]

\beq Q_{jm}=P_{mj}\Bigm/\sum_{i=1}^3 P_{mi},\eeq
The observer's final ignorance level, after receiving output $m$, is
given by the Shannon entropy,

\beq H_m=-\sum_{j=1}^3 Q_{jm}\,\ln{Q_{jm}}. \eeq
This need not be, but often is, less than the initial entropy,

\beq H_{\rm init}=-\sum_{j=1}^3 p_j\,\ln{p_j},\eeq
so that some information has been gained, even though the result is
inconclusive.\bigskip

\medskip\noindent{\bf 5. Higher dimensional space}\medskip

\noindent Finally, let us briefly outline how the above results can be
generalized to $N$ signals ($N>3$). Consider the $N$-th order matrix
formed by the components of all the input vectors, in any basis. Instead
of the triple product $[\bu_1\bu_2\bu_3]$, we now have the determinant
of that matrix. Vector products $\bv_j$ such as in Eq.~(\ref{v}) become
outer products of any $N-1$ signal states. Their components, in any
basis, are the appropriate cofactors in the above determinant.  The
argument leading to Eq.~(\ref{A0}) remains essentially the same, and we
now obtain a $(N-1)$-dimensional hyper\-surface in the $N$-dimensional
$k$-space. It is plausible that this hyper\-surface is convex in the
first orthant (i.e., hyper-octant) in $k$-space. A formal proof of this
conjecture is a straightforward but tedious exercise in differential
geometry (perhaps a more clever proof can be found).  Optimization then
proceeds as in Sect.~3, by considering a family of parallel
hyper\-planes $\sum B_jk_j=X$.

There are now many possibilities of partial answers. For example, if the
signal states $\bu_j$ can be divided into two (or more) mutually
orthogonal subspaces, it is possible, in a first step, to determine
unambiguously the subspace to which each signal belongs. Then, a second
step is to try to identify individual non-orthogonal signals within a
given subspace.

An interesting problem is how to utilize the resulting mixed
information, with some of the signals fully identified, and others only
partly identified. For example, if we have two mutually ortho\-gonal
subspaces, and in each one two non-ortho\-gonal states, an individual
state encodes two bits, but a subspace is still worth  one bit, plus
some amount of mutual (probabilistic) information. Further investigation
is needed to clarify this issue.\clearpage

\bigskip\noindent{\bf Acknowledgments}\medskip

DRT was supported by a grant from the Technion Graduate School. Work by
AP was supported by the Gerard Swope Fund, and the Fund for
Encouragement of Research.\bigskip 

\noindent{\bf References}

\frenchspacing \begin{enumerate}
\item Bennett C H 1992 {\it Phys. Rev. Letters\/} {\bf 68} 3121
\item Davies E B 1978 {\it IEEE Trans. Inform. Theory\/} {\bf IT-24} 239
\item Helstrom C W 1976 {\it Quantum Detection and Estimation Theory\/}
(New York: Academic Press) Chapt 4
\item Dieks D 1988 {\it Physics Letters A\/} {\bf 126} 303 
\item Peres A 1988 {\it Physics Letters A\/} {\bf 128} 29
\item Jaeger G and Shimony A 1995 {\it Physics Letters A\/} {\bf 197} 83
\item Chefles A 1998 {\it Physics Letters A\/} {\bf 239} 339
\item Peres A 1993 {\it Quantum Theory: Concepts and Methods\/}
(Dordrecht: Kluwer) pp 282--285

\vfill\noindent CAPTION OF FIGURE\bigskip

\noindent{\bf Figure 1.} Domain of positivity of $\bA_0$.

\end{enumerate}
\end{document}